\documentclass{emulateapj}
\usepackage{apjfonts,color}
\shortauthors{Morrissey et al.}
\begin{document}
\title{The On-Orbit Performance of the Galaxy Evolution Explorer}

\author{
Patrick Morrissey,\altaffilmark{1}\email{patrick@srl.caltech.edu}
David Schiminovich,\altaffilmark{1}
Tom A. Barlow,\altaffilmark{1}
D. Christopher Martin,\altaffilmark{1}
Brian Blakkolb,\altaffilmark{2}
Tim Conrow,\altaffilmark{1} 
Brian Cooke,\altaffilmark{2}
Kerry Erickson,\altaffilmark{2}
James Fanson,\altaffilmark{2}
Peter G. Friedman,\altaffilmark{1}
Robert Grange,\altaffilmark{3}
Patrick N. Jelinsky,\altaffilmark{5}
Siu-Chun Lee,\altaffilmark{2}
Dankai Liu,\altaffilmark{2}
Alan Mazer,\altaffilmark{2}
Ryan McLean,\altaffilmark{1}
Bruno Milliard,\altaffilmark{3}
David Randall,\altaffilmark{2}
Wes Schmitigal,\altaffilmark{2}
Amit Sen,\altaffilmark{2}
Oswald H. W. Siegmund,\altaffilmark{5}
Frank Surber,\altaffilmark{2}
Arthur Vaughan,\altaffilmark{2}
Maurice Viton,\altaffilmark{3}
Barry Y. Welsh,\altaffilmark{5}
Luciana Bianchi,\altaffilmark{4}
Yong-Ik Byun,\altaffilmark{10} 
Jose Donas,\altaffilmark{3}
Karl Forster,\altaffilmark{1}
Timothy M. Heckman,\altaffilmark{4}
Young-Wook Lee,\altaffilmark{2}
Barry F. Madore,\altaffilmark{6,7}
Roger F. Malina,\altaffilmark{3}
Susan G. Neff,\altaffilmark{8}
R. Michael Rich,\altaffilmark{9}
Todd Small,\altaffilmark{1}
Alex S. Szalay\altaffilmark{4} and
Ted K. Wyder\altaffilmark{1}}

\altaffiltext{1}{California Institute of Technology, MS 405-47, 1200 East
California Boulevard, Pasadena, CA 91125}
\altaffiltext{2}{Jet Propulsion Laboratory, California Institute of Technology, 
4800 Oak Grove Drive, Pasadena, CA 91109}
\altaffiltext{3}{Laboratoire d'Astrophysique de Marseille, BP 8, Traverse
du Siphon, 13376 Marseille Cedex 12, France}
\altaffiltext{4}{Department of Physics and Astronomy, The Johns Hopkins
University, 3400 N. Charles St., Baltimore, MD 21218}
\altaffiltext{5}{Space Sciences Laboratory, University of California at
Berkeley, 7 Gauss Way, Berkeley, CA 94720}
\altaffiltext{6}{Observatories of the Carnegie Institution of Washington,
813 Santa Barbara St., Pasadena, CA 91101}
\altaffiltext{7}{NASA/IPAC Extragalactic Database, California Institute
of Technology, Mail Code 100-22, 770 S. Wilson Ave., Pasadena, CA 91125}
\altaffiltext{8}{Laboratory for Astronomy and Solar Physics, NASA Goddard
Space Flight Center, Greenbelt, MD 20771}
\altaffiltext{9}{Department of Physics and Astronomy, University of
California, Los Angeles, CA 90095}
\altaffiltext{10}{Center for Space Astrophysics, Yonsei University, Seoul
120-749, Korea}

\begin{abstract}
We report the first year on-orbit performance results for the Galaxy Evolution Explorer (GALEX), a NASA Small Explorer that is performing a survey of the sky in two ultraviolet bands.  The instrument comprises a 50~cm diameter modified Ritchey-Chr\'{e}tien telescope with a $1.25^{\circ}$ field of view, selectable imaging and objective grism spectroscopic modes, and an innovative optical system with a thin-film multilayer dichroic beam splitter that enables simultaneous imaging by a pair of photon counting, microchannel plate, delay line readout detectors.  Initial measurements demonstrate that GALEX is performing well, meeting its requirements for resolution, efficiency, astrometry, bandpass definition and survey sensitivity.
\end{abstract}

\keywords{space vehicles: instruments --- surveys --- telescopes --- ultraviolet: general}

\section{Introduction}

\notetoeditor{This paper is submitted for consideration in the GALEX special issue of the Astrophysical Journal Letters.}

The Galaxy Evolution Explorer (GALEX) is a NASA Small Explorer mission currently performing an all-sky ultraviolet survey in two bands.  GALEX was launched on an Orbital Sciences Corporation
 Pegasus rocket on 2003 April 28 at 12:00~UT from the Kennedy Space Center into a circular, 700~km, 29$^{\circ}$ inclination orbit.  The instrument is designed to image a very wide 1.25$^{\circ}$\ field of view with 4~--~6\arcsec\ resolution and sensitivity down to m$_{\rm AB}\sim 25$\ in the deepest modes.  

GALEX makes science observations on the night side of each orbit during ``eclipses'' that are typically in the range of 1500~--~1800~s.  In the first year of operations, we have observed over 4000 square degrees of sky and accumulated nearly a terabyte of science data.  The science mission is described in a companion paper by \citet{Martin2004}.

\section{Instrument Overview}

\begin{figure}
\epsscale{1.2}
\plotone{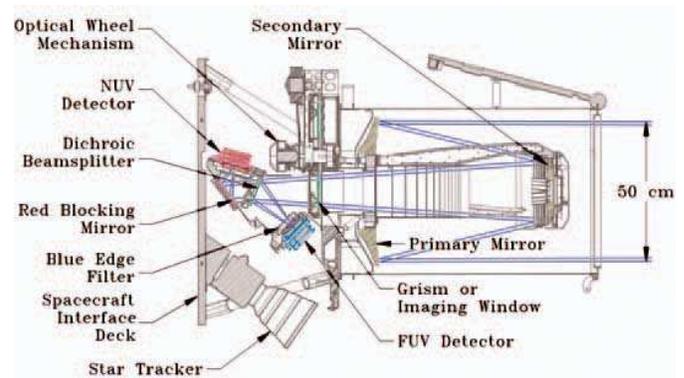}
\caption{A cross section of the instrument portion of GALEX.  The optical path is outlined in blue.  Overall dimensions of the view shown are $1.5\times1$~m.  Solar panels and a separate compartment for spacecraft bus electronics (supplied by Orbital) are not included.\label{xsection}}
\end{figure}

The instrument, shown in Fig.~\ref{xsection}, comprises a 50~cm diameter modified Ritchey-Chr\'{e}tien telescope with two photon-counting, microchannel plate (MCP),
 delay line readout detectors that image the sky simultaneously in the near and far ultraviolet (NUV and FUV).
An optical wheel mechanism behind the primary mirror allows
selection of an imaging window, dispersive grism, or opaque shutter.
Downstream of the optical wheel, a multilayer-coated dichroic
beam splitter enables
simultaneous imaging in the two bands.  A blue edge multilayer filter blocks
geocoronal Lyman-$\alpha$~$\lambda$1216~\AA\ and O\textsc{I}~$\lambda$1304~\AA\ 
emission from the FUV channel, while a red
blocking fold mirror reduces contamination from zodiacal light ($\lambda>3000$~\AA) in
the NUV band.  Details of the optical design are provided in
Tab.~\ref{prescription}.
 
\begin{deluxetable}{ll}
\tabletypesize{\scriptsize}
\tablecaption{The GALEX Optical Prescription.}
\tablewidth{0pt}
\tablehead{\colhead{Parameter} & \colhead{Value}}
\startdata
\textbf{Telescope:}\\
\hspace{2em}Type:              & Ritchey-Chr\'{e}tien\\
\hspace{2em}Coatings:          & MgF$_{\rm 2}$-coated Al\\
\hspace{2em}Primary diameter:  & 500 mm\\
\hspace{2em}Secondary diameter:& 230 mm\\
\hspace{2em}Focal length:      & 2998 mm\\
\hspace{2em}Focal ratio:       & 6\\
\hspace{2em}Plate scale:       & 68.80\arcsec-mm$^{-1}$\\

\textbf{Grism (CaF$_{\rm 2}$):}\\
\hspace{2em}Width (inscribed diameter): & 124 mm\\
\hspace{2em}Thickness (center):& 5.9 mm\\
\hspace{2em}Wedge:             & 1.37$^{\circ}$\\
\hspace{2em}Blaze Angle:       & 2.33$^{\circ}$\\
\hspace{2em}Ruling:            & 75 lines-mm$^{-1}$\\

\textbf{Imaging Window (CaF$_{\rm 2}$):}\\
\hspace{2em}Diameter:          & 124 mm\\
\hspace{2em}Thickness:         & 6 mm\\

\textbf{Dichroic (Fused Silica):}\\
\hspace{2em}Diameter:         & 110 mm\\
\hspace{2em}Thickness:        & 4 mm\\
\hspace{2em}Wedge:            & 0.119$^{\circ}$\\

\textbf{Blue Edge Filter (MgF$_{\rm 2}$):}\\
\hspace{2em}Diameter:         & 74 mm\\
\hspace{2em}Thickness:        & 2.5~mm\\

\textbf{Red Blocking Mirror (Fused Silica):}\\
\hspace{2em}Diameter:         & 110 mm\\
\enddata
\label{prescription}
\end{deluxetable}

The pair of detectors at the heart of the instrument are each
completely sealed tubes with no active pumping \citep{Jelinsky2003}.  Each contains a photocathode, a stack of 3 MCPs, and a delay-line anode with outputs at each of 4 corners.  The 68~mm active area (75~mm diameter) Z-stack MCPs operate at a high voltage (HV) gain of approximately $1.5\times 10^7$ (-6200~V FUV, -5200~V NUV), producing pulses for each incoming photon that are divided by the anode and measured at each corner to determine the photon position, which is computed by the detector electronics and stored in a Solid State Recorder (SSR) for later download.   The FUV detector has a MgF$_{\rm 2}$ window and an opaque CsI photocathode deposited directly on the front microchannel plate, while the NUV detector has a proximity-focused semi-transparent Cs$_{\rm 2}$Te photocathode deposited on the vacuum side of its fused silica window, 300~microns from the MCP surface.

\section{Instrument Performance}

The instrument was calibrated in thermal vacuum during a
month-long period in mid-2001; these data form a performance baseline for
flight measurements.  Resolution, background,
flat field, throughput, linearity, spectral resolution and spatial distortion
were measured in several sets of thermal conditions.  In some areas the ground
calibration provides data not easily measurable in flight, particularly the
imaging-mode bandpass and grism-mode dispersion.  In flight, white dwarf standard stars tie the
calibration to \textit{HST} and establish the photometric zero point in each band.

To date, GALEX has observed six \textit{HST} white dwarf standard stars \citep{Bohlin2001}, which span five magnitudes in UV intensity.  Details of these observations are listed in Tab.~\ref{standards}, with key performance indicators presented in Tab.~\ref{performance} and in the remainder of this section.

\begin{deluxetable}{lllll}
\tabletypesize{\scriptsize}
\tablecaption{\textit{HST} White Dwarf Standards observed by GALEX.\label{standards}}
\tablewidth{0pt}
\tablehead{
\colhead{Star} 
& \colhead{m$_{\rm FUV}$\tablenotemark{a}} 
& \colhead{m$_{\rm NUV}$\tablenotemark{a}} 
& \colhead{$\alpha$(2000)\tablenotemark{b}} 
& \colhead{$\delta$(2000)\tablenotemark{b}}}
\startdata
HZ21        & 12.55 & 13.13 & 12$^h$ 14$^m$ \phantom{0}1.6$^s$ 
            & 32$^{\circ}$ 58\arcmin\ 11.3\arcsec\\ 
HZ43        & 10.75 & 11.36 & 13\phantom{$^h$} 16\phantom{$^m$} \phantom{0}5.8\phantom{$^s$}  
            & 29\phantom{$^{\circ}$} \phantom{0}5\phantom{\arcmin}\ 55.0\phantom{\arcsec}\\
HZ44        & 10.02 & 10.27 & 13\phantom{$^h$} 23\phantom{$^m$} 35.4\phantom{$^s$}  
            & 36\phantom{$^{\circ}$} \phantom{0}8\phantom{\arcmin}\ \phantom{0}3.5\phantom{\arcsec}\\
BD$+33^{\circ}2642$ & 10.51 & 10.47 & 15\phantom{$^h$} 51\phantom{$^m$} 59.86\phantom{$^s$} 
            & 32\phantom{$^{\circ}$} 56\phantom{\arcmin}\ 54.8\phantom{\arcsec}\\
LDS749B     & 15.57 & 14.71 & 21\phantom{$^h$} 32\phantom{$^m$} 16.3\phantom{$^s$}  
            & \phantom{0}0\phantom{$^{\circ}$} 15\phantom{\arcmin}\ 14.4\phantom{\arcsec}\\
G93-48      & 12.14 & 12.39 & 21\phantom{$^h$} 52\phantom{$^m$} 25.3\phantom{$^s$}  
            & \phantom{0}2\phantom{$^{\circ}$} 23\phantom{\arcmin}\ 17.9\phantom{\arcsec}\\
\enddata
\tablenotetext{a}{Magnitudes shown are predictions based on the GALEX bandpass and publicly available reference spectra from the \textit{HST} CALSPEC database at http://www.stsci.edu/instruments/observatory/cdbs/calspec.html.}
\tablenotetext{b}{Coordinates are GALEX-measured and include proper-motion through the 2003-2004 epoch.  One exception is BD$+33^{\circ}2642$, which was only observed in grism mode; the coordinate provided is from \citet{Bohlin2001}.}
\end{deluxetable}

\begin{deluxetable}{lcc}
\tabletypesize{\scriptsize}
\tablecaption{Summary of measured performance parameters for GALEX.\label{performance}}
\tablewidth{0pt}
\tablehead{\colhead{Item} & \colhead{FUV Band} & \colhead{NUV Band}}
\startdata
Bandwidth:\tablenotemark{a}    & 1344~--~1786~\AA             & 1771~--~2831~\AA \\
Effective wavelength ($\lambda_{\rm eff}$):\tablenotemark{b}
                               & 1528~\AA                     & 2271~\AA\\
Field of view:                 & 1.28$^{\circ}$               & 1.24$^{\circ}$\\
Peak effective area:           & 36.8 cm$^2$ at 1480~\AA      & 61.7 cm$^2$ at 2200~\AA\\
Zero point ($m_{\rm 0}$):      & 18.82                        & 20.08\\
Image resolution (FWHM):       & 4.5\arcsec                   & 6.0\arcsec\\
Spectral resolution ($\lambda/\Delta\lambda$):
                               & 200                          & 90 \\
Detector background (typical):\\
\hspace{2em}Total:             & 78 c-s$^{-1}$                & 193 c-s$^{-1}$\\
\hspace{2em}Diffuse:           & 0.66 c-s$^{-1}$-cm$^{-2}$    & 1.82 c-s$^{-1}$-cm$^{-2}$\\
\hspace{2em}Hotspots:          & 47 c-s$^{-1}$                & 107 c-s$^{-1}$\\
Sky background (typical):\tablenotemark{c} 
                               & 2000 c-s$^{-1}$              & 20000 c-s$^{-1}$\\
Limiting magnitude ($5\sigma$):\tablenotemark{d}\\
\hspace{2em}AIS (100~s):       & 19.9                         & 20.8\\
\hspace{2em}MIS (1500~s):      & 22.6                         & 22.7\\
\hspace{2em}DIS (30000~s):     & 24.8                         & 24.4\\             
Linearity:\\
\hspace{2em}Global (10\% roll off):     & \multicolumn{2}{c}{18000 c-s$^{-1}$} \\
\hspace{2em}Global (50\% roll off):     & \multicolumn{2}{c}{91000 c-s$^{-1}$} \\
\hspace{2em}Local (10\% roll off):\tablenotemark{e}
                               & 89 c-s$^{-1}$                & 471 c-s$^{-1}$\\
Pipeline image format: & \multicolumn{2}{c}{$3840\times3840$ elements with 1.5\arcsec\ pixels}\\
\enddata
\tablenotetext{a}{The bandpass is defined by wavelengths with effective area at least 10\% of the peak.}
\tablenotetext{b}{\citet{Fukugita1996}, Equation~3.}
\tablenotetext{c}{These correspond to 569 and 1189~photons-s$^{-1}$-cm$^{-2}$-sr$^{-1}$-\AA$^{-1}$, respectively.}
\tablenotetext{d}{Approximate All Sky (AIS), Medium (MIS) and Deep (DIS) Imaging Survey depths.}
\tablenotetext{e}{These are worst-case values for point sources.}
\end{deluxetable}

\subsection{Astrometry}

GALEX observations are dithered in a 1.5\arcmin\ spiral in order to smooth out the effects of small
scale detector distortions and flat field variations; the pipeline system must
determine the spacecraft and detector transformation for each photon in the list to form sharp images.  We have measured the astrometric performance against 
Tycho-2 catalog \citep{Hog2000} reference stars to refine the plate solution of the instrument, with the result that 80\% of stars in the central 1$^{\circ}$ of the detector area are currently found within 2.8\arcsec\ of their expected location in the FUV and 1.5\arcsec\ in the NUV.  We expect these figures to decrease for the whole detector area as the calibration is improved with flight data.

\subsection{Background}

The intrinsic detector background is negligible, dominated by isolated hotspots that result mainly from microscopic defects at the MCP surface.  The remaining diffuse component is enhanced by a factor of 2 over the very low values observed on the ground.  These results are typical for an MCP detector placed in the low earth orbit space radiation environment and are consistent with little or no contribution from optics phosphorescence. 
More significant is the background from the sky, which varies during each eclipse by about 20\%.  This signal is dominated by diffuse galactic light in the FUV and by zodiacal light in the NUV.  These are orders of magnitude greater than the detector background and, along with the PSF, define the limiting magnitude of the deep surveys.

Other sources of background include: bright star glints in the NUV images near the edge of the detector window; large out-of-focus pupil images resulting from reflections in the imaging window and dichroic beam splitter; an arc-minute diameter skirt around bright NUV sources (at the 0.5\% level) resulting from the proximity-focused cathode; and photoemission from the quantum efficiency (QE) enhancing grid wires on the FUV detector window that appear as linear features around bright sources.  Work is being done to flag these locations automatically in the pipeline software, but they amount to a very small fraction of the data ($\ll 1$\%).\footnote{Examples of GALEX data are available at http://www.galex.caltech.edu.}

\subsection{Photometry}

GALEX uses the AB magnitude system of \citet{Oke1983} with the FUV and NUV magnitudes defined as follows:
\[m_{\rm UV} = m_{\rm 0} - 2.5\log f_{\rm UV}\]
Here, $f_{\rm UV}$ is the dead-time-corrected count rate for a given source \textit{divided by the flat field map,} which is of order 1, and $m_{\rm 0}$ is the zero point that corresponds to the AB magnitude of a 1~count-s$^{-1}$ (cps) flat-field-corrected detection.  
There are independent relative response functions and zero points for each band.  The white dwarf observations
show remarkable agreement with ground test data from 2001, with only a 10\% (0.1 magnitude) decrease in the NUV band, and essentially perfect agreement in FUV. 
The apparent lack of degradation (and therefore contamination) is attributable to a careful contamination control program in conjunction with the non-cryogenic MCP-based design.  

Repeat observations demonstrate that the current low resolution flat field (based on ground calibration data) is performing at a level better than $\pm 0.1$ magnitudes RMS, however there are some subtleties that it does not capture.  Chief among these is probably the shadow caused by the QE-enhancing grid wires in the FUV detector.  Some work has already been done with stacks of data from deep observations at different roll angles, and these show the grid wires and also several other details quite clearly.  It is expected that incorporation of these features will improve photometry, with the most significant gains for sources with short exposures that do not execute a complete dither pattern. 

\subsubsection{Count Rate Linearity and Bright Star Constraints}

There are essentially two sources of photometric non-linearity in the
instrument: global dead time resulting from the finite time required for the
electronics to assemble photon lists, and local dead time
resulting from the MCP-limited current supply to small regions around bright sources.  
Global dead time is a linear
function of the input count rate.  It is easily measured using an on-board
``STIM'' pulser, which electronically stimulates each detector anode with a
steady, low rate stream of electronic pulses that are imaged off the field of
view.  Since the real rate of STIM pulses is accurately known, the measured
rate is used by the pipeline to scale the effective exposure and thus correct the
global dead time for all sources in the field simultaneously.  This correction
is typically about 10\% in NUV and negligible in FUV, however it can become
quite significant ($\sim 50$\%) for the brightest fields. 

\begin{figure}
\epsscale{1.2}
\plotone{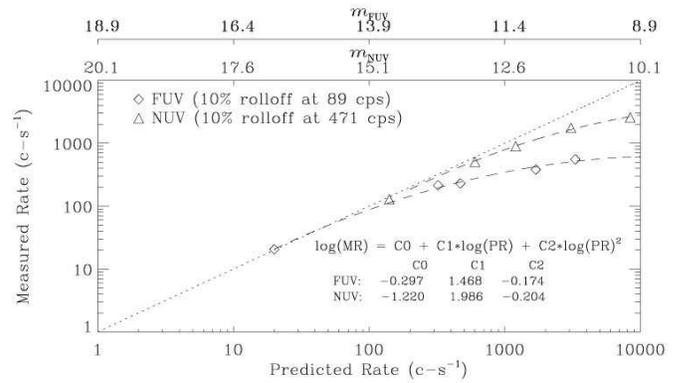}
\caption{Local photometric non-linearity measured with white dwarf
standards.  Empirical fits are overplotted.  The function parameters $PR$ and $MR$ correspond to the predicted and measured source rates, respectively.\label{locallin}}
\end{figure}

Local dead time is difficult to correct with high accuracy, since it depends on the source distribution.  It affects the measured count rate and shape of individual bright sources.  We have used our standard stars to estimate the local dead time in each band as
shown in Fig.~\ref{locallin}.  The NUV detector is more robust
to bright sources because it is proximity-focused and thus presents a larger
image (with lower count density) to the MCP.      
Note that since local photometric non-linearity is strongly source-size dependent,
the saturation shown in Fig.~\ref{locallin} for stars is a worst-case scenario.

GALEX currently adheres to a bright star constraint of 5000~cps, which corresponds to approximately 10th magnitude.  This significant survey constraint is being enforced in the early stage of the mission to avoid affecting the MCP gain.

\subsection{Resolution}

The width of the point spread function (PSF) is defined by numerous contributors, including the optics and pipeline reconstruction, however the detectors add a large share, particularly in the proximity-focused NUV channel.  To verify the fundamental instrument performance from on-orbit data, some bright stars were analyzed individually, outside the pipeline.  These results show consistent or better performance than what was measured during ground tests.  Currently the pipeline system is achieving about 6.5\arcsec\ full width at half maximum (FWHM) in the FUV and 7.2\arcsec\ in the NUV with a broad distribution in both cases.  We expect to be able to improve these each significantly (to approximately the instrument-limited values listed in Tab.~\ref{performance}) with improvements in calibration -- particularly distortion map refinements -- and aspect correction algorithms.

\subsection{Spectroscopy}

\begin{figure}
\epsscale{1.2}
\plotone{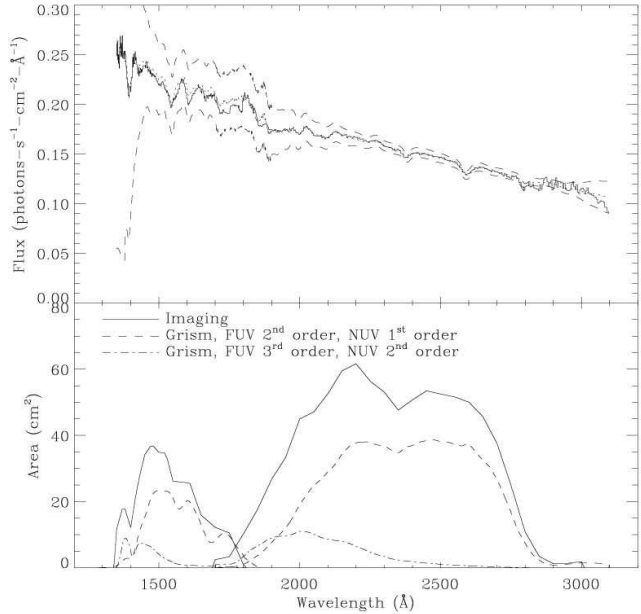}
\caption{\textit{Upper panel:} (Solid line) A composite of 42 $\sim$240~s GALEX spectra of the \textit{HST} white dwarf standard star BD$+33^{\circ}2642$.  (Dotted line) The \textit{HST} reference spectrum.  (Dashed lines) The range of $1\sigma$ variations among the 42 spectra.  \textit{Lower panel:} (Solid lines) The ground-measured imaging-mode effective area.  (Dashed lines) The flight-measured grism-mode effective area.  (Dot-Dashed lines) The additional effective area in the grism 2nd (NUV) and 3rd (FUV) orders.\label{bdp33d2642}}
\end{figure}

The spectroscopy mode utilizes a CaF$_{\rm 2}$ grism in the converging beam of the telescope to form simultaneous spectra of all sources in the field.  In a typical image, many spectra will overlap and multiple observations with different grism angles can be combined to eliminate confusion.  Some early results are presented in Fig.~\ref{bdp33d2642} for the white dwarf standard BD$+33^{\circ}2642$.

We combined 42 spectra taken with different roll angles and at different field positions to compute the grism response in the primary orders (1st for NUV and 2nd for FUV).  There is significant response in the next higher order of each band that will eventually allow some broadening of the grism bandpass.  The grism effective area for the FUV band is about 15\% greater on average than the response computed on the ground, while the
NUV band is about 10\% less on average.  After applying our flat-field correction, the computed grism response in the FUV varies by about 10\% across the detector, and for the NUV varies by about 3\%.

The FWHM resolution derived from absorption lines in the dispersion direction is between
5~--~6\arcsec\ ($\sim 8$\AA) in the FUV and 6~--~7\arcsec\ ($\sim 26$\AA) in the NUV, although it may degrade towards the ends of the bandpass.   The dispersion function agrees with ground-based calibrations to within a fraction of a resolution element.  The offset of the spectra from the direct-mode image, which is determined by correlating extracted stars in any given field with UV stellar spectra templates, is accurate to about 1\arcsec.

\section{Mission Efficiency}

Since launch, GALEX has had numerous detector events that are correlated with space weather.  In addition to occasional HV overcurrents, which are quickly quenched by the high speed on-board fault protection, the FUV background has jumped tens of thousands of
cps several times, and remained elevated until after a period of rest with the HV turned off.
The first major example of this occurred on 2003 October 28, closely correlated with an historically large (X17) solar flare.  Flight spare laboratory tests demonstrate that charging of the front window surface enhances the detector background, and that discharging the window makes the background dissipate completely.  We expect our operational efficiency, currently averaging about 70\%, to improve as we benefit from several flight software patches that have addressed the space weather, window charging and other details.

\section{Summary}

We have reported the GALEX instrument performance during the first year on orbit.
The satellite continues to perform well, and since it has no consumables, 
has the potential to continue to observe for many years.  
Near term goals include improving the flight calibration to help the pipeline achieve instrument-limited performance, and gradually opening the detector count rate limits based on laboratory experiments to reduce bright-star constraints on survey coverage.

\acknowledgments

We gratefully acknowledge NASA's support for construction, operation,
and science analysis for the GALEX mission,
developed in cooperation with the Centre National d'Etudes Spatiales
of France and the Korean Ministry of 
Science and Technology. The grating, window, and aspheric corrector were supplied by France.
We acknowledge the dedicated
team of engineers, technicians, and administrative staff from JPL/Caltech, 
Orbital, University of California, Berkeley, 
Laboratoire d'Astrophysique de Marseille, 
and the other institutions who made this mission possible.

\end{document}